\documentclass{PoS}

\long \def \blockcomment #1\endcomment{}

\def\ie{{\it i.e.}}

\def\tb{\tilde{b}}
\def\vx{{\vec x}}
\def\vy{{\vec y}}
\def\vk{{\vec k}}

\def\vp{{\vec p}}

\def\vj{{\vec j}}

\def\hT{\hat{T}}
\def\hF{\hat{\Phi}}
\def\hP{\hat{\Pi}}
\def\hH{\hat{H}}

\def\cu{{\cal U}}
\def\ch{{\cal H}}

\def\beq{\begin{equation}}
\def\eeq{\end{equation}}
\def\bqry{\begin{eqnarray}}
\def\eqry{\end{eqnarray}}

\title{Transfer Matrix for Partially Quenched QCD}

\ShortTitle{Transfer Matrix for Partially Quenched QCD}

\author{Claude Bernard
\\
        Department of Physics, Washington University, St. Louis, MO 63130, USA\\
        E-mail: \email{cb@lump.wustl.edu}}

\author{\speaker{Maarten Golterman}\\
        Department of Physics and Astronomy, San Francisco State University,
        San Francisco, CA 94132, USA\\
        E-mail: \email{maarten@stars.sfsu.edu}}

\abstract{We construct the transfer matrix for the ghost sector of partially quenched
QCD.  This transfer matrix is not hermitian, but we show that it is still bounded.
We thus expect that all euclidean correlation functions will decay exponentially
with distance (up to possible powers), and
demonstrate that this is indeed the case for free ghost quarks.}

\FullConference{The XXVIII International Symposium on Lattice Field Theory, Lattice2010\\
		June 14-19, 2010\\
		Villasimius, Italy}

\begin{document}

\section{Introduction}
The use of ``partially quenched''  QCD (PQQCD), in which valence and sea quark
masses are varied independently, has become common practice in large-scale
computations of hadronic quantities from QCD using lattice methods.  This usefulness
arises from the following observations:
\begin{itemize}
\item[1.]  Sea quark masses are, by definition, the quark masses that go into the
generation of dynamical gauge field configurations on which hadronic quantities
are evaluated.  Valence quark masses are, by definition, the quark masses that
go into the operators that are evaluated on these gauge configurations.  Such
operators are contractions of quark propagators, which depend on the
gauge field configurations.  For many applications, it is numerically less expensive
to generate quark propagators than gauge field configurations.   Given limited
resources it can thus be advantageous to generate data for many values of
the valence quark masses, using configurations depending on only a limited
number of sea quark mass values.
\item[2.]
PQQCD with a given number of sea quarks contains full QCD with the same
set of sea quarks \cite{BG1994}.    It follows that the low-energy constants
(LECs) of the low-energy effective theory are those of the real world, because
by construction, these LECs are independent of the quark masses
\cite{ShSh2000}.\footnote{They do depend on the number of sea quark flavors.}  
For the case of unquenched QCD, this low-energy theory is
chiral perturbation theory (ChPT), which  provides a systematic expansion in
(light) quark masses, with coefficients characterized by these LECs.
\item[3.]
On the lattice, PQQCD can be generalized to ``mixed-action'' QCD, in which
not only valence and sea quark masses are independently chosen, but also
the discretization of the Dirac operator is chosen independently in the sea
and valence sectors of the theory.   This is a generalization in the sense that the
continuum limit of such a lattice theory is described by PQQCD \cite{BRS2003}.  
\end{itemize}

In order to apply these ideas, one needs the correct effective theory for
PQQCD, \ie, one needs the partially quenched version of ChPT (PQChPT)
\cite{MG2009}.  While it is relatively straightforward to extend ChPT to the
partially quenched case, it is less clear than in the unquenched case whether
indeed PQChPT is the correct effective theory for PQQCD.  This is because
PQQCD violates a number of basic properties of a healthy quantum field
theory on which arguments that ChPT is the correct effective theory for 
unquenched QCD rely.  This follows from the definition of PQQCD as
a euclidean path integral that includes an integral over ghost quarks, which
have the same quantum numbers as the valence quarks, but the opposite (and
thus ``wrong'') statistics.  

For many field theories, the path integral for a euclidean theory 
can be expressed in terms of a transfer matrix.  If the field theory is healthy, this transfer matrix is hermitian and
bounded, and contact with the hamiltonian formulation of the theory can be made.\footnote{Of
course. quantum field theories exist which are believed to be healthy, but for which
no transfer matrix can easily be constructed.}
But if the theory contains fields with the wrong statistics, it is not clear what properties
of such a construction survive.

In Ref.~\cite{SW1979}, the validity of ChPT as a systematic low-energy effective
theory for the Goldstone sector of QCD was conjectured to follow from the
basic properties of a healthy quantum field theory: analyticity, unitarity, cluster
decomposition, and symmetry.  Stated differently, the $S$-matrix calculated with
the most general local lagrangian consistent with a certain symmetry group
was conjectured to be the most general possible $S$-matrix consistent with
these basic properties.  This was then used as a starting point for the systematic
development of ChPT as an expansion of $S$-matrix elements in terms of the pion momenta, following a well-defined power-counting scheme.
The reliance of this argument on unitarity, though, appears to be a fundamental difficulty
in trying to extend it to the partially quenched case, which is certainly not unitary.

An alternative justification for ChPT as the low-energy effective theory for QCD
was presented in Ref.~\cite{HL1994}. The argument detailed there was based
on locality and clustering of the underlying theory of unquenched QCD, as well as its 
symmetries.
Locality and clustering
guarantee the existence of vertices in the effective theory that
are independent of the correlation functions in which they appear, and,
consequently, the existence of a loop expansion. This approach seems more
fruitful for PQQCD than that of Ref.~\cite{SW1979}.  By construction, PQQCD is
local.   It is less clear that it also satisfies the cluster property, but
numerical evidence suggests that that is indeed the case, because euclidean
correlation functions are observed to decay exponentially in
distance (up to possible nonstandard powers, which come from the characteristic
double poles \cite{BG1994,ShSh2001} in the theory).

The
first requirement for the chiral theory of PQQCD is, of course, the
dynamical breaking of chiral symmetry in the theory with massless quarks.
The Goldstone bosons associated with this breaking provide the low-energy
degrees of freedom for the low-energy effective theory.  It was argued in Ref.~\cite{ShSh2001} that 
chiral symmetry must also be broken in PQQCD.  The argument is that this 
must happen in the sea
sector of that theory, because it is identical to unquenched QCD. This follows, since, by
construction, the valence (and ghost) quarks are not part of the dynamics.  
Goldstone bosons corresponding to pions made only out of sea quarks thus
have to be present in the partially quenched theory. 
Furthermore, the vectorlike
partially quenched symmetries that relate sea and valence quarks then imply that there also have to be Goldstone excitations in the
valence and ghost sectors, thus providing the necessary degrees of freedom
for the construction of PQChPT.\footnote{These vectorlike symmetries are not
spontaneously broken \cite{VW1983}.}

The key outstanding issue for the justification of PQChPT as the correct chiral theory
for PQQCD therefore seems to be the question of whether PQQCD obeys the clustering
property.
Here we begin an investigation into the clustering property of PQQCD by
considering the construction of a transfer matrix
for PQQCD.  We start with the ghost sector, which is the ultimate
source for the nonstandard features of the partially quenched theory.

\section{Staggered ghosts}
PQQCD contains three types of quarks: sea quarks, valence quarks, and 
ghost quarks, which have the same masses as the valence quarks, but opposite
statistics.  Because of this quark content, the complete fermion determinant of the
partially quenched theory is just that coming from the sea sector, while
the valence and ghost determinants cancel each other.

Since ghost quarks violate spin-statistics, we expect that obstructions to
the existence of a well-behaved transfer matrix would originate from this
sector of the theory.  We therefore consider a theory of {\rm bosonic} staggered
quarks, in an arbitrary gauge field background.\footnote{We use staggered
quarks here because the nonperturbative definition of PQQCD with staggered
quarks is straightforward.  For other types of quarks, see Ref.~\cite{MG2009}
and refs. therein, in particular Ref.~\cite{GSS2005}.}  If we choose
\bqry
\label{action}
S&=&\sum_{x,y}\chi^\dagger(x)D(x,y)\chi(y)+\sum_x m\chi^\dagger(x)\chi(x)\\
&=&\sum_x\left\{\frac{1}{2}\sum_\mu\eta_\mu(x)\left(\chi^\dagger(x)U_\mu(x)\chi(x+\mu)
-\chi^\dagger(x+\mu)U^\dagger_\mu(x)\chi(x)\right)+m\chi^\dagger(x)\chi(x)\right\}\ ,
\nonumber
\eqry
with
\beq
\label{phases}
\eta_\mu(x)=(-1)^{x_1+\dots+x_{\mu-1}}\ ,
\eeq
then
\beq
\label{Z}
Z=\int D\chi^\dagger D\chi\;\mbox{exp}\left(-S\right)
\eeq
is convergent for $m>0$, because the staggered Dirac operator $D$ is 
anti-hermitian.  The field $\chi$ is bosonic, \ie, it is a $c$-number valued staggered quark
field.  Of course,
$S$ exhibits the same species doubling as the normal staggered quark action, and therefore we employ a two time-slice
method for constructing the transfer matrix representation of $Z$, guided by the
construction in Ref.~\cite{Smit}.

We split $\chi$ into its real and imaginary parts by defining 
$\chi(x) = \eta_4(x)\phi_1(x) + i\phi_2(x)$, and then identify
\bqry
\label{ident}
t=2k:\ \ \ \ \ \ &\phi_1({\vec x},t)=\Phi_{1,k}({\vec x})\ ,\ \ \ \ \ \phi_2({\vec x},t)=-\Phi_{2,k}({\vec x})\ ,\\
t=2k+1:\ \ \ &\!\!\!\!\!\!\!\!\phi_1({\vec x},t)=\Pi_{2,k}({\vec x})\ ,\ \ \ \ \ \phi_2({\vec x},t)=\Pi_{1,k}({\vec x})\nonumber\ .
\eqry
Then, if the extension of the (periodic) lattice in the time direction is $T$ (which we take to be even), the partition function can be written as the trace of the $T/2$-th power of a
transfer matrix,
\beq
\label{transf}
Z({\cal U})=\mbox{Tr}\left(\prod_{k=1}^{T/2}\hT_k({\cal U})\right)\ ,
\eeq
in which 
\beq
\label{tmatrix}
\hT_k(\cu)=e^{-\hF_1\ch_-[\cu(2(k+1)]\hF_2}\,e^{-\hP_2\ch_+[\cu(2k+1)]\hP_1}\ .
\eeq
Here $\ch_\pm$ are defined through
\bqry
\label{chdef}
\Psi_1\ch_\pm[\cu(t)]\Psi_2&=&\sum_\vx\Biggl\{\pm\sum_j i\eta'_j(\vx)\left(\Psi_1(\vx)\;\mbox{Re}\;
U_j(\vx,t)\Psi_2(\vx+\vj)+(1\leftrightarrow 2)\right)\\
&&\hspace{0.75cm}-\sum_j i\eta_j(\vx)\left(\Psi_1(\vx)\;\mbox{Im}\;
U_j(\vx,t)\Psi_1(\vx+\vj)-(1\to 2)\right)\nonumber\\
&&\hspace{0.75cm}+m\left(\Psi_1(\vx)^2+\Psi_2(\vx)^2\right)\Biggr\}\ ,\nonumber\\
\eta'_j(\vx)&=&\eta_j(\vx)\eta_4(\vx)\ ,\nonumber
\eqry
and the hermitian operators $\hF_{1,2}$ and $\hP_{1,2}$ satisfy the commutation rules
\beq
\label{commrules}
[\hF_a(\vx),\hP_b(\vy)]=i\delta(\vx-\vy)\delta_{ab}\ .
\eeq
{}From $\hat T$, we can define a hamiltonian in the limit in which the temporal
lattice spacing goes to zero; with
\bqry
\label{H}
{\hH}[{\cu(t)}]&=&\lim_{a_t\to 0}-\log{\hT}_k({\cal U})/(2a_t)\ ,\\
{\hH}[{\cu(t)}]&=&{\hH}_1+i{\hH}_2[{\cu(t)}]\ ,\nonumber\\
\hH_1&\!=\!&\frac{1}{2}\;m\sum_{\vx}\left(\hP_1^2(\vx)+\hP_2^2(\vx)+\hF_1^2(\vx)+\hF_2^2(\vx)\right)\ ,\nonumber\\
\hH_2[\cu(t)]&\!=\!&\frac{1}{2}\sum_{\vx,j}\Biggl\{
\eta'_j(\vx)\left(\hP_2(\vx)\;\mbox{Re}\;U_j(\vx,t)\hP_1(\vx+\vj)-
\hF_1(\vx)\;\mbox{Re}\;U_j(\vx,t)\hF_2(\vx+\vj)\right)
\nonumber\\
&&\hspace{0.8cm}-\eta_j(\vx)\left(-\hP_1(\vx)\;\mbox{Im}\;U_j(\vx,t)\hP_1(\vx+\vj)+
\hF_1(\vx)\;\mbox{Im}\;U_j(\vx,t)\hF_1(\vx+\vj)
\right)\nonumber\\
&&\hspace{0.8cm}+(1\leftrightarrow 2)\Biggr\}\ .
\nonumber
\eqry
Both $\hH_1$ and $\hH_2$ are hermitian, and do not commute.  Therefore,
$\hH$ is not hermitian, and not normal.  It follows that the transfer matrix is not
hermitian and not positive definite.

Despite these (nonsurprising) conclusions, it is possible to make progress.
Our transfer matrix factorizes as
\beq
\label{factorize}
\hT_k(\cu)=\hT_1(\cu)\hT_2(\cu)\ ,
\eeq
in which $\hT_1$ is the exponential operator with $\ch_-$ and $\hT_2$ is that with
$\ch_+$.  Both $\hT_1$ and $\hT_2$ are normal and bounded. Since the
hamiltonians $\ch_-$ and $\ch_+$ have a positive real part (the parts proportional
to $m$):
\beq
\label{bound12}
\parallel \hT_{1,2}\parallel\,\le 1\ ,
\eeq
which implies
\beq
\label{boundT}
\parallel \hT\parallel\,\le\parallel \hT_1\parallel\parallel \hT_2\parallel\,\le 1\ ,
\eeq
which establishes that all eigenvalues of $\hT$ have an absolute value smaller
than one.\footnote{This product inequality is satisfied if we use the euclidean norm,
which, for a matrix $A$, is defined as the square-root of the largest eigenvalue of
$A^\dagger A$.} It follows that correlation functions
in this theory decay exponentially with distance if the eigenvalue $\lambda_0$
with maximal $|\lambda_0|$ is unique.
Of course, the construction needs
to be extended to the complete partially quenched theory, but we do not expect
any difficulties with the other building blocks of a complete transfer matrix.

\blockcomment
Let us assume that $\hH$ has a complete set of left and right eigenstates
$|L_\lambda\rangle$ and $|R_\lambda\rangle$ with eigenvalue $\lambda$.\footnote{
This is true for a trivial gauge field background, and we expected it to be true for all gauge fields except for a possible
set of measure zero.}  Then
\beq
\label{deriv}
2\langle R_\lambda|\hH_1|R_\lambda\rangle=
\langle R_\lambda|(\hH[\cu(t)]+\hH^\dagger[\cu(t)])|R_\lambda\rangle
=(\lambda+\lambda^*)\langle R_\lambda|R_\lambda\rangle\ ,
\eeq
hence
\beq
\label{bound}
\mbox{Re}\;\lambda=\frac{\langle R_\lambda|\hH_1|R_\lambda\rangle}
{\langle R_\lambda|R_\lambda\rangle}\ge 0\ .
\eeq
We find that, despite the nonhermiticity of $\hH$, the real part of its eigenvalues
are bounded from below, and this leads us to expect that correlation functions
in this theory decay exponentially with distance (if there is a mass gap).  We note
that  it is possible to obtain a bound directly on the transfer matrix, $|\hT|\le 1$,
which, of course, implies the bound~\ref{bound}.  
\endcomment

\section{Trivial gauge field background}
For a trivial gauge field background, it is straightforward to diagonalize $\hH$.
In terms of canonical creation and annihilation operators, introduced through
\bqry
\label{cran}
\Phi_1(\vx)&=&\int\frac{d^3k}{(2\pi)^3}\;\frac{1}{\sqrt{2}}\left(a_1(\vk)+a_1^\dagger(-\vk)\right)e^{i\vk\cdot\vx}\ ,\\
\Pi_1(\vx)&=&\int\frac{d^3k}{(2\pi)^3}\;\frac{-i}{\sqrt{2}}\left(a_1(\vk)-a_1^\dagger(-\vk)\right)e^{i\vk\cdot\vx}\ ,\nonumber\\
\Phi_2(\vx)&=&\int\frac{d^3k}{(2\pi)^3}\;\frac{-i}{\sqrt{2}}\left(a_2(-\vk)-a_2^\dagger(\vk)\right)e^{i\vk\cdot\vx}\ ,\nonumber\\
\Pi_2(\vx)&=&\int\frac{d^3k}{(2\pi)^3}\;\frac{-1}{\sqrt{2}}\left(a_2(-\vk)+a_2^\dagger(\vk)\right)e^{i\vk\cdot\vx}\ ,
\eqry
we find that
\beq
\label{free}
\hH=\int_\vp\left\{m\left(a^\dagger_1(\vp)a_1(\vp)+a^\dagger_2(\vp)a_2(\vp)\right)
+i\sum_j\sin(p_j)\left(a_1(\vp)\alpha^j a_2(\vp)-a^\dagger_2(\vp)\alpha^j a^\dagger_1(\vp)\right)
\right\}\ ,
\eeq
with $-\pi/2 < p_j \le \pi/2$.
The $\alpha^j$ are anti-hermitian $8\times 8$ Dirac matrices because
there are $2^3$ spatial doublers. (The two doublers in 
in the time direction appear explicitly.)  Now $\sum_j\sin(p_j)\alpha^j$ can be diagonalized, with eigenvalues
\beq
\label{evs}
\pm is(p)\equiv\pm i\sqrt{\sum_j\sin^2(p_j)}\ ,
\eeq
making $\hH$ a sum of terms of the form
\beq
\label{Hpart}
h(\vp)=m\left(a^\dagger_1(\vp)a_1(\vp)+a^\dagger_2(\vp)a_2(\vp)\right)
\pm s(p)\left(a_1(\vp)a_2(\vp)-a^\dagger_2(\vp)a^\dagger_1(\vp)\right)\ .
\eeq
For each $\vp$, this can be diagonalized with a generalized Bogoliubov transformation
\cite{Swanson}:
\bqry
\label{Boogie}
b_1&=&\cos\theta\,a_1-\sin\theta\,a_2^\dagger\ ,\\
b_2&=&\cos\theta\,a_2-\sin\theta\,a_1^\dagger\ ,\nonumber\\
\tb_1&=&\cos\theta\,a^\dagger_1+\sin\theta\,a_2\ ,\nonumber\\
\tb_2&=&\cos\theta\,a^\dagger_2+\sin\theta\,a_1\ ,\nonumber
\eqry
where we note that $\tb_i$ is {\em not} the hermitian conjugate of $b_i$.
With $\theta=\frac{1}{2}\tan^{-1}(s/m)$ this yields
\beq
\label{hbo}
h=E\left(\tb_1 b_1+\tb_2 b_2\right)\ ,\qquad E=\sqrt{m^2+s^2}\ .
\eeq
The operators $b_i$ and $\tb_i$ are annihilation and creation operators, 
and indeed, $\hH$ has complete sets of left and right eigenstates.
The form of the eigenvalues $E$ are not a surprise: the ghost determinant should
cancel the valence determinant, so we expect the eigenvalues for the
ghost hamiltonian to match those of the valence hamiltonian.

We can also calculate correlation functions in the theory with a trivial
background.  The two-point functions in a theory with hamiltonian~(\ref{free})
are
\bqry
\label{twopoint}
\langle a_i(t)a^\dagger_j(0)\rangle&=&\delta_{ij}\,\frac{E+m}{2E}\,e^{-Et}\ ,\\
\langle a^\dagger_i(t)a_j(0)\rangle&=&-\delta_{ij}\,\frac{E-m}{2E}\,e^{-Et}\ ,\nonumber\\
\langle a_i(t)a_j(0)\rangle&=&
-\langle a^\dagger_i(t)a^\dagger_j(0)\rangle=\delta_{i+j,3}\,\frac{s}{2E}\,e^{-Et}\ .\nonumber
\eqry
While these correlation functions exhibit the expected exponential decay, they also clearly show a violation of unitarity.  In a healthy field theory, by inserting a complete set
of states between the operator at time $t$ and the one at time $0$, one would
conclude that $\langle a^\dagger_i(t)a_j(0)\rangle$ has to be nonnegative.  Here this is
not the case, because the vacuum bra $\langle 0|$ defined by $\langle 0|\tb_i=0$ is not the conjugate of the vacuum
ket $|0\rangle$ defined by $b_i|0\rangle=0$ since $\tb_i$ is not the hermitian
conjugate of $b_i$.
The results~(\ref{twopoint}) also follow directly from a 
path integral for this hamiltonian, without first performing the Bogoliubov transformation.

\section{Discussion}

In this talk, we took the first, and most important, step toward the construction of
a transfer matrix for PQQCD.  We considered
PQQCD with staggered fermions, because it is straightforward to define the ghost sector
in that case.  The ghost quarks are then ``staggered ghosts,'' \ie, staggered quarks with bosonic statistics, and we constructed the transfer matrix for staggered ghosts in an
arbitrary gauge field background.  The rest of the construction remains to be done,
but we anticipate that adding the sea and valence parts, as well as
the gauge part, to the transfer matrix will be straightforward.

We then proved that the ghost transfer matrix in an arbitrary gauge field background
is bounded in the sense that the absolute value of its largest eigenvalue is bounded
by one.  At nonzero quark mass, correlation functions of ghost quarks are thus expected
to decay exponentially, even if they do not satisfy the positivity properties of a healthy
quantum field theory.  We would expect this to carry over to PQQCD as long as that
theory has a nonvanishing mass gap.   In the case of PQQCD, the meaning of this
would be that there is a unique eigenvalue $\lambda_0$ of the
complete transfer matrix of the partially quenched theory with $|\lambda_0|$ 
maximal, while all other eigenvalues $\lambda$ satisfy $|\lambda|<|\lambda_0|$.

We then checked this in the free ghost theory, in the limit of vanishing temporal
lattice spacing.  Indeed, we find that there is a nonzero
mass gap, set by the quark mass $m$, and all two-point functions decay exponentially.
This supports the conjecture that the argument of Ref.~\cite{HL1994} for the 
validity of ChPT can be carried over to the partially quenched case.   Already in the
free theory we find violations of unitarity in two-point functions, so it is clear that also
PQQCD, and therefore PQChPT, will suffer from the same disease.  However,
PQChPT would still be the correct effective theory for PQQCD at low energy,
because the underlying theory is local, and still satisfies cluster decomposition,
despite its nonhermitian transfer matrix.

We thank Michael Ogilvie for helpful discussions.

\end{document}